\documentstyle[aaspp4]{article}
\begin{document}

\title{The IMF and its Evolution}
\author{Bruce Elmegreen}
\affil{IBM T.J. Watson Research Center, P.O. Box 218, Yorktown Hts., NY
10598, USA}

\begin{abstract}
Observations of the stellar initial mass function are reviewed. The IMF
is flat, or possibly declining, below several tenths of a Solar mass,
and declining above this mass in a power law with a slope of about
$-1.35$ on a log-log plot.  The flattening at low mass is evidence for
a characteristic mass in star formation, which, according to recent
theory, may be either the minimum stellar mass for the onset of
deuterium burning, or the thermal Jeans mass.  The first of these
masses should not vary with environment as much as the second, so any
observed variations in the mass of the flattened part are important for
understanding star formation.  Starburst galaxies may be an example
where the characteristic mass is larger than it is locally, but this
old observation has been challenged lately.  A steeper high-mass slope
in the extreme field studied by Massey et al. may be the result of
cloud destruction and the termination of star formation by ionization,
with a normal IMF in each separate cluster.  The lack of a density
dependence in cluster IMFs suggests that star and protostar
interactions play little role in star formation or the IMF. This is unlike
the case for binaries and disks, which do show an environmental
influence, and all are consistent with the observed stellar density in
clusters, which is high enough to promote interactions between binaries
and disks, but not individual stars.  These considerations,
along with indirect observations of the IMF in the early Universe,
suggest that the IMF does not vary much in its basic form over position
and time, but that shifts in the characteristic mass might occur in
regions with extremely high or low star formation activity, 
or perhaps light-to-mass ratio, with the
characteristic mass, star formation efficiency, and gas consumption
rate all following the light-to-mass ratio.
\end{abstract}

to appear in The evolution of Galaxies on Cosmological Timescales,
ed. J.E. Beckman and T.J. Mahoney, ASP Conference Series, 
San Francisco, 1999, in press. 

\section{Introduction}

Observations of clusters and associations suggest an {\it average}
stellar initial mass function (IMF) that is approximately a power law
like the Salpeter (1955) function, with a slope of $x\sim1.35$ on a
$\log n-\log M$ plot, and a flattening below $\sim0.35$ M$_\odot$. This
IMF appears in clusters and whole galaxies, for all galactic
populations, and even in the intergalactic medium (Sect. 2.1, 2.3, 2.4).
However, there are still fluctuations in the slope of the power-law by
$\pm0.5$ from cluster to cluster (Scalo 1998), and there are other
curious variations too, like 
a steeper slope in the field (Sect. 2.2), the 
mass of the most massive star
increasing with cloud mass (Sect. 3), the formation of
massive stars relatively late and near the centers of clusters (Sect.
3), and the greater proportion of massive stars in
starburst galaxies (Sect. 2.5). Considering the robust nature
of the IMF, any theory for its origin should be able to
reproduce both the average shape and the variations around it with a
minimum of free parameters and a minimal dependence on the physical
properties of the star-forming clouds.

Another important mass function for star formation is the distribution
of cloud and clump masses. This differs from the stellar function
in both slope ($x\sim0.5-0.8$ for clouds) and range
($M_{cloud}\sim10^{-4}-10^7$ M$_\odot$; Heithausen et al. 1998; Dickey
\& Garwood 1989), leading one to wonder why stars form with a steeper
mass distribution than their clumps. There must be a preferential selection
of lower clump masses for stars, and a cutoff at some minimum star mass.

There are tantalizing indications that we may be able to understand
the IMF without fully understanding the origin of either
the cloud structure or the processes involved with 
individual stars. {\it Given} the observed structures of clouds, we can 
imagine how star formation processes might select
pieces of this structure in a certain order and end up with
the observed IMF and all of its variations. If such clump selection is the
correct explanation for the IMF, then it presumably works because 
most of the star mass is determined by the gas mass 
immediately available to it during the protostar phase, and because
the IMF is an average over many different processes, with each
losing its unique contribution when the mass distribution is averaged over
a cluster.

Numerical simulations of such sampling demonstrate this point
by reproducing essentially all of the observations of the IMF and 
its systematic and stochastic 
variations without any free parameters or physical input other than
a single characteristic mass for the minimum clump that can make a
star. These models obtain (Elmegreen 1997, 1999a): (1) the correct
power-law slope and turnover shape of the IMF, with the correct turnover
mass, (2) the tendency for the most massive star in a cluster to
increase with cloud mass, (3) the shift in the peak or turnover mass
for starburst regions
without a change in the power-law slope, (4) the
delayed formation of massive stars in a cluster, (5) the fluctuations in
the slope of the power-law part from cluster to cluster (which result
from sampling statistics), and (6) the tendency for the most massive
stars in a cluster to concentrate toward the cluster center. The only
input to the model is the hierarchical (and fractal) distribution of cloud structure, and
the only assumption is that pieces of this hierarchy make stars at a rate that
scales with the square root of the local density, which is the rate at which
essentially all of the physical processes involved with the onset of
star formation operate, including self-gravity, magnetic diffusion, clump
collisions, and turbulence dissipation, given the molecular cloud scaling laws. 

The hypotheses that IMF theories may be simplified by the gross
averaging of star formation processes during the build up of a
cluster, and by the intimate connection between its power-law slope and
cloud structure, also help to explain why its power-law slope 
is so similar from
region to region, even in different environments and at different times.
The point is that the cloud and star formation details may not
matter much for the IMF, and that power-law cloud structures are more-or-less
universal, perhaps as a result of pervasive turbulence.

In the next section we review the observations of the IMF and
some of the implications of these observations in an attempt
to sort out what is physically significant and compelling for a theory.
Other reviews can be
found in the conference proceedings {\it The Stellar Initial Mass
Function}, edited by Gilmore, Parry \& Ryan for ASP Press in 1998. 
A review that compares various theories with the constraints from
observations is in Elmegreen (1999b).

\section{Observations of the IMF and Implications for the Theory}
\subsection{The Salpeter Slope in Clusters and Galaxies}

The IMF at intermediate to high mass can be written $n(M)d\log M\propto
M^{-x}d\log M$ for slope $x$ on a $\log-\log$ plot. For most clusters,
$x$ is in the range 1--1.5. Salpeter (1955) suggested $x\sim1.35$, which
is about the average of the values observed today. The most dependable
values come from a mixture of photometry and spectroscopy of star
clusters. IMFs based on photometry alone are generally steeper than
$x\sim1.35$ because of an ambiguity in mass for high mass stars (see
discussion in Massey 1998). 

\begin{table}
\caption{Observations of a Salpeter IMF with $x=1-1.5$}
\begin{tabular}{ll}
\tableline
Star counts in clusters& Elson, Fall, \& Freeman 1989; Massey \& \\
& Thompson 1991; Vallenari et al. 1992; Elson 1992\\
& Massey \& Johnson 1993; Phelps \& Janes 1993\\
& Hillenbrand et al. 1993; Drissen et al. 1993\\
& Parker \& Garmany 1993; Ninkov et al. 1995\\
& Massey et al. 1995; Chiosi et al. 1995; Banks, \\
& Dodd, \& Sullivan 1995; Will, Bomans, \\
& \& de Boer 1995; Will et al. 1995, 1997; Deeg\\
& \& Ninkov 1996; Hunter et al. 1996a,b, 1997\\
& Massey \& Hunter 1998\\
Star counts in the field&Scalo 1986 ($x\sim1-1.5$ for 
intermediate mass stars)\\
H$\alpha$ equiv. widths and &Greggio et al. 1993; Kennicutt, Tamblyn,\\
galaxy photometry&\& Congdon 1994; Marconi et al.  1995\\
& Bresolin \& Kennicutt 1997; Holtzman, et al.\\
& 1997; Grillmair et al.  1998\\
Elemental abundances in&\\
\hspace{0.2in}local ISM:&Tsujimoto et al. 1997\\
\hspace{0.2in}galactic stellar halo:&Nissen et al. 1994\\
\hspace{0.2in}QSO damped Ly$\alpha$:&Lu et al. 1996\\
\hspace{0.2in}Ly$\alpha$ forest:&Wyse 1998\\
\hspace{0.2in}intracluster medium:&Renzini et al. 1993, Wyse 1997, 1998\\
& (but see Loewenstein \& Mushotzky 1996)\\
\hspace{0.2in}elliptical galaxies:&Wyse 1998\\
Galaxy evolution models&Sommer-Larson 1996; Lilly et al. 1996\\
\tableline
\tableline
\end{tabular}
\end{table}

Table 1 summarizes the recent observations that obtain $x\sim1-1.5$ in
various regions. This ``Salpeter'' slope is found 
by star counts in local clusters,
integrated light from whole galaxies, elemental abundances, and
galaxy evolution models. Steeper values of $x\sim1.5-2$ are found in
samples of local field stars or in the low density parts of some
clusters (Table 2). Shallower values are found at low mass, where the
IMF flattens to nearly zero slope on a $\log-\log$ plot (Table 3).
Shifts either in the peak or in the slope, favoring higher masses,
have been found in starburst galaxies (Table 4). 

The observations in these 
tables suggest that the IMF varies a lot, but in fact
most of the functions that deviate from the turned-over Salpeter slope are based on
indirect measurements that contain questionable assumptions. For
example, the slope determined for the local field tends to get steep
only at high mass, and the increased value depends on an assumed
recent star formation history and an assumed scale height variation with
mass and age. The local field is also more populated by low mass
stars than high mass stars because low mass stars live longer and drift
further from their sites of star formation than high mass stars.

The low density regions of clusters show a steeper IMF too because of an
excess of low mass stars, but this is probably related to the greater
concentration of high mass stars in cluster cores, as discussed more
in Section 3; the
overall cluster can still have a flattened-Salpeter 
IMF. The {\it Hipparcos}
results quoted by Brown (1998) were based on photometry, rather than
spectra, and are typically steep for photometry. Massey et al. (1995) has
shown how such IMF values become shallower, like the Salpeter function,
when spectra are considered for the determination of stellar mass.

\subsection{A Steep IMF Slope in The Extreme Field}

The most extreme deviation for an IMF measurement is in the remote field
regions of the LMC and Milky Way (Table 2). 
These are regions defined by Massey et
al. (1995) to be further than 30 pc from the boundaries of catalogued OB
associations. Here the slope at high mass has been measured to be around
$x\sim4$. Evidently something very unusual is happening. There are
several ways to explain this, if it turns out to be true. One way
has a normal ($x\sim1.35$) IMF in every
individual region of star formation, and a steeper IMF in the
composite of many regions. This difference between
cluster and intergrated IMFs illustrates an important point about cloud
destruction, so we discuss it in some detail here (see also Elmegreen
1999a).

\begin{table}
\caption{Observations of the IMF with $x=1.5-2$ or greater}
\begin{tabular}{ll}
\tableline
Local field stars&Miller \& Scalo 1979; Garmany, Conti,\\
& \& Chiosi 1982; Humphreys \& McElroy 1984\\
& Blaha \& Humphreys 1989; Basu \& Rana 1992\\
& Kroupa, Tout, \& Gilmore 1993; Scalo 1986 \\
&($x=1.5-2$ for high mass stars)\\
Local OB associations&({\it review} of Hipparcos results: Brown 1998)\\
LMC clusters in regions&J.K. Hill et al. 1994; R.S. Hill et al. 1995\\
low young-star density& \\
Unclustered embedded&Ali \& DePoy 1995\\
stars in Orion&\\
Extreme field stars&Massey et al. 1995\\
in the LMC ($x\sim4$)&\\
\tableline
\tableline
\end{tabular}
\end{table}

In a large region there will in general be many separate clouds that
form stars, and these clouds will have some mass function
$n(M_{c})dM_{c}\propto M_{c}^{-\gamma} dM_{c}$ for
$\gamma\sim1.5-2$. {\it If intermediate and high mass stars} destroy their
clouds because of ionization, and as a result, {\it halt the star formation processes
inside them}, then more massive clouds will require
more massive stars before star formation ends. This leads to a situation where a
lot of low mass clouds make primarily low mass stars, with a normal IMF,
and where a few high
mass clouds make both low mass and high mass stars, 
also with a normal IMF. But, since there are
more low mass clouds, the composite region will have a lot more low mass
stars in proportion to high mass stars
than is given by each cluster IMF. 
It follows that even if the IMF inside each region of star
formation is the same, a Salpeter IMF for example, the composite IMF
from many clouds will be steeper than this. 

Consider a specific example. Suppose the IMF in each region of star
formation has a certain slope $x$, and the largest mass of a star,
$M_L$, required to destroy a cloud scales with cloud mass $M_c$ as
$M_{L}\propto M_{c}^{\alpha}$ for $\alpha>0$. Then 
the composite IMF from all of 	the clouds combined
will have a slope $x_{comp}=\left(\gamma-1\right)/\alpha,$ which is
{\it independent of the IMF slope in each individual cluster}.

To evaluate this composite slope, we take $\gamma=2$ for a hierarchical
cloud system (Fleck 1996; Elmegreen \& Falgarone 1996), 
and $\alpha=5/16$ for cluster destruction with a largest
stellar mass $M_{L}$. This value of $\alpha$ comes from the
mass-luminosity relation of ionizing radiation, which scales as
$L\propto M^4$ for luminosity $L$ and stellar mass $M$ (Vacca, Garmany,
\& Shull 1996). A whole cluster's ionizing luminosity can be evaluated
from the expression $\int_0^{M_L} L(M)n(M)dM$ for maximum mass $M_L$ and
IMF $n(M)dM=xM_L^xM^{-1-x}dM$. This cluster luminosity scales with $M_L^4$ too.
The constant term in the IMF, $xM_L^x$, gives one star at a maximum mass
$M_L$ from the expression $\int_{M_L}^\infty n(M)dM=1$. The luminosity
required to destroy a cloud is the binding energy divided by the cloud
crossing time, which is $\left(GM_c^2/R\right)
\left(GM_c/R^3\right)^{1/2}\propto M_c^{5/4}$, using the Larson (1981)
scaling laws for molecular clouds. Setting the luminosity of a cluster,
$\propto M_L^4$, equal to the power required to destroy a cloud,
$\propto M_c^{5/4}$, then gives $\alpha=5/16$ in the expression
$M_L\propto M_c^\alpha$.

With $\gamma=2$ and $\alpha=5/16$, the slope of the composite
IMF is $x_{comp}=\left(\gamma-1\right)/\alpha=16/5\sim3.2$.
The value observed by Massey et al. (1995) is $\sim4$, which
is pretty close to this theoretical result, given the
uncertainties in the $M-L$ relation and other assumptions, and
with the observations.

It is important to note that the extreme field IMF found by Massey et al.
(1995) is not representative of galaxies in general. Integrated light
and elemental abundances give an average IMF for whole galaxies that has
the same slope at intermediate and high mass as individual clusters,
namely, the Salpeter value of $x\sim1.35$. This simple fact implies that
{\it massive stars cannot generally halt star formation in their
clouds}. If they did, then the composite IMF for a whole galaxy would be
significantly steeper than the individual IMF in each cluster. Massive
stars may {\it destroy} their clouds, in the sense that they push the
gas around, but they cannot generally halt star formation in them except
possibly in the extreme field. The extreme field could differ from the
environment in OB associations because of a much lower pressure in the
extreme field. A low pressure could conceivably lead to more efficient
cloud ionization and the cessation of star formation in even the dense
clumps.

The requirement that the composite IMF be equal to the cluster IMF also
means that $\alpha=1/x$ in the above analysis (with $\gamma=2$, as
required for a hierarchical gas distribution). This is just what is
expected for random star formation, where the largest stellar mass
increases with cloud mass simply because of random sampling from the
IMF. That is, the largest stellar mass satisfies $\int_{M_L}^\infty
n(M)dM=1$, as discussed above, and this gives a constant of
proportionality $n_0=xM_L^x$ in the expression $n(M)=n_0M^{-1-x}$. Thus
the total number of stars scales with $M_L^x$. If the efficiency is
about constant with cloud mass (and the smallest mass star is much 
less massive than $M_L$), then this total number scales about with
the cloud mass, giving $M_L^x\propto M_c$, or $\alpha=1/x$.

There are other explanations for the steep IMF in the extreme
field. Star forming regions are typically much larger than 30 pc, often
extending in a coherent fashion up to several hundred parsecs (Efremov
1995), so the 30 pc limit in the definition of the extreme field
may allow some normal cluster, association, or star-complex
members to be included.  In that case, the steep slope in the outer
regions of a cluster may occur for the same reasons as the shallow
slope in the inner region, i.e., segregation of the most massive stars
towards the center.

In summary, the general form of the 
IMF is probably invariant among clouds of
different masses, giving a maximum stellar mass that increases
with cloud mass as the power $1/x=1/1.35$ as a result of random
sampling (i.e., more massive clouds sample further out into the
high mass tail of the IMF).  This explains the similarity
between the composite IMF of whole galaxies and the IMFs
of individual clusters. However, in the extreme field, 
where conditions like ambient pressure are very different
than in OB associations, star-forming clouds could be more
quickly and easily destroyed by ionization from stars, 
and in this case, the maximum stellar mass could increase much
more slowly with cloud mass, as the power $1/4$ instead of
$1/1.35$.  As a result, the composite IMF can be much
steeper than the individual IMFs in each cluster.
Alternatively, the extreme field IMF could be sampling only the
low mass members of an extended cluster whose other members
are more centrally located. 

\subsection{An IMF that is Independent of Cluster Density}

One of the most startling aspects of the observed IMF is that it is
virtually {\it invariant from cluster to cluster}, aside from likely
statistical fluctuations (Elmegreen 1999a), and this relative invariance
spans a range of a factor of 200 in cluster 
density (Hunter et al. 1997; Massey \& Hunter 1998) and
a factor of 10 in metal abundance (Freedman 1985; Massey, Johnson \&
DeGioia-Eastwood 1995). 

The density independence means that {\it the IMF is probably not the
result of protostar, star, or clump interactions}. If it were, then dense
regions, which should have more of these interactions, would differ from
low density regions, where there are few or no interactions. The IMF is also
{\it not likely to result from accretion of cloud material during
stellar orbital motion}. Stars in denser regions orbit in a shorter time
and have more gas to accrete. Neither is the IMF or any part of it
from the {\it coalescence} of stars (i.e., massive stars are not
formed from the coalescence of low mass stars or protostars).

This lack of a density dependence for individual stars in the IMF
contrasts with the situation for binary stars and disks. The binary
fraction is smaller in denser regions, and protostellar disks are
smaller too (see review in Elmegreen et al. 1999). The protostellar
binary fraction is lower in both the Trapezium cluster (Petr et al.
1998) and the Pleiades cluster (Bouvier et al. 1997) than it is in the
Tau-Aur region, by a factor of $\sim3$. Also, the peak in the separation
distribution for binaries is smaller (90 AU) in the part of the Sco-Cen
association that contains early type stars than it is (215 AU) in the
part of the Sco-Cen association that contains no early type stars
(Brandner \& K\"ohler 1998).

The cluster environment also apparently affects disks. Mundy et al.
(1995) suggested that massive disks are relatively rare in the Trapezium
cluster, and N\"urnberger et al. (1997) found that protostellar disk
mass decreases with stellar age in the Lupus young cluster, but not in
the Tau-Aur region, which is less dense. When massive stars are present,
as in the Trapezium cluster, uv radiation can photoionize the
neighboring disks (Johnstone et al. 1998).

These observations make sense in terms of the relative interaction rates
for stars, binaries, and disks (Elmegreen et al. 1999). 
The size of a typical embedded cluster is
$\sim0.1$ pc, and the number of stars is several hundred. This makes the
stellar density on the order of $10^3-10^4$ stars pc$^{-3}$. For
example, in the Trapezium cluster, the stellar density is $\sim5000$
stars pc$^{-3}$ (Prosser et al. 1994) or higher (McCaughrean \& Stauffer
1994), and in Mon R2 it is $\sim9000$ stars pc$^{-3}$ (Carpenter et al.
1997). A stellar density of $10^3$ M$_\odot$ pc$^{-3}$ corresponds to an
H$_2$ density of $\sim10^4$ cm$^{-3}$. Molecular cores with densities of
$10^5$ cm$^{-3}$ or higher (e.g., Lada 1992) can easily make clusters
this dense, because star formation efficiencies are typically 10\%-40\%
(e.g., see Greene \& Young 1992; Megeath et al. 1996; Tapia et al.
1996).  

The density of $n_{star}=10^3$ stars pc$^{-3}$ in a cloud core of size
$R_{core}\sim0.2$ pc implies that objects with this density will collide
with each other in one crossing time
if their cross section is
$\sigma\sim \left(n_{star}R_{core}\right)^{-1} \sim0.005$ pc$^2$, which
corresponds to a physical size of
$\sim6500\left(R_{core}[pc]n_{star}/10^3\right)^{-1/2}$ AU. This is the
size of protostellar disks and long-period binary stars.  Thus
disks and 
binaries should be affected by interactions in the cluster environment, 
but not individual stars or the IMF. 

\subsection{The Flattening at Low Mass: a Characteristic Mass for Stars}

The IMF flattens on a $\log-\log$ plot at stellar masses of
around and below $0.3$ M$_\odot$. Table 3 summarizes the observations. The mass at
which this flattening occurs is observed to vary a bit from region to
region, particularly in clusters (i.e., the mass at the peak in NGC 6231
is 2.5 M$_\odot$, much higher than normal; Sung, Bessell, \& Lee 1998), but such
variations could be the result of mass segregation in the sense that high
mass stars are often concentrated towards cluster cores (see Sect.
3). There is even evidence for a turnover in the IMF at masses less
than 0.3 M$_\odot$ for several regions, but this is uncertain because
the stars at the low mass end are usually close to the limit of
detection.

\begin{table}
\caption{Observations at Low Mass}
\begin{tabular}{ll}
\tableline
Flat, $x\sim0-0.5$:&\\
Clusters&Reid 1987; Kroupa, Tout \& Gilmore 1990, 1991\\
& Hubbard, Burrows, \& Lunine 1990; Zuckerman\\
&\& Becklin 1992; Kroupa, Gilmore,\& Tout 1992\\
& Tinney, Mould, \& Reid 1992; Tinney 1993\\
& Laughlin \& Bodenheimer 1993; Comeron et al.\\
& 1993; Jarrett, Dickman \& Herbst 1994; Paresce,\\
& de Marchi, \& Romaniello 1995; Strom, Kepner,\\
& \& Strom 1995; Pound \& Blitz 1995; Williams,\\
& Rieke \& Stauffer 1995; Williams, et al. 1995\\
& Kroupa 1995a; Comeron, Rieke, \& Rieke 1996\\
& Meusinger et al. 1996; Macintosh, et al. 1996\\
& Festin 1997; Hillenbrand 1997; Luhman\\
& \& Rieke 1998; Reid 1998\\
Binary star mass ratios&Tout 1991\\
Lack of Brown Dwarfs&Zuckerman \& Becklin 1992; Reid \& Gazis 1997\\
& Reid 1998\\
Globular clusters, halo,&de Marchi \& Paresce 1997; Chabrier \& Mera 1997\\ 
and bulge stars&Holtzman et al. 1998\\
Rise at lower mass&Tinney 1993; Mera et al. 1996; Zapatero Osorio\\
&et al. 1997\\
Turnover at lower mass&Reid \& Gazis 1997; Hillenbrand 1997; Reid 1998\\
&Sung et al. 1998; Nota et al. 1998; King et al. 1998\\
&Gould, et al. 1997\\
\tableline
\tableline
\end{tabular}
\end{table}

The importance of the IMF flattening is that this is the only characteristic
scale known for the star formation process. Molecular clouds and their
pieces have a power law mass distribution from sub-stellar masses to the
masses of clouds as big as the galactic scale height. There is
essentially no characteristic scale for clouds. 
The mass distributions for open clusters and perhaps even primordial
globular clusters are power laws too, with about the same
slope as for clouds (Elmegreen \& Efremov 1997; see review in 
Elmegreen et al. 1999). 
The rest of the
IMF is a power law too. But the IMF does have
a characteristic scale at the low mass end, where it flattens
at about 0.3 M$_\odot$.

The existence of such a characteristic mass is an important clue to the
mechanism of star formation. For example, we know now that
the characteristic mass is not the Jeans mass at an
optical depth of unity, as formerly suggested,
because this mass is too small, $\sim 10^{-3}$ M$_\odot$ (e.g., Rees
1976). The two most promising suggestions for the origin of the
characteristic mass are: (1) self-limitation of accretion by
protostellar winds triggered at the deuterium-burning mass 
(Nakano, Hasegawa, \& Norman 1995; Adams \& Fatuzzo 1996), and (2)
the inability of a cloud piece smaller than the
thermal Jeans mass to become self-gravitating and collapse to a star,
given the temperature and pressure of a molecular cloud core (Larson 1992;
Elmegreen 1997).

The first of these limits would seem to be relatively independent of
environment, while the second should scale with $T^2/P^{1/2}$ for cloud
temperature $T$ and cloud-core pressure $P$. Both values are about the
same locally, where $T\sim10$K and $P\sim10^6$ k$_{\rm B}$ cm$^{-3}$,
and since $T^2$ and $P^{1/2}$ tend to vary together with galactocentric
radius and star formation activity (Elmegreen 1997, 1999b), the two
masses should remain the same in most normal regions.

To check the theoretical predictions, we should look for places where
$T^2/P^{1/2}$ deviates a lot from its local value. If the mass at the
peak of the IMF, or where the IMF flattens, varies from region to region
along with the quantity $T^2/P^{1/2}$, then the second model would be
preferred; if the peak mass does not, then the first model is better.
For example, Larson (1998) suggested that the peak in the IMF was
shifted towards higher masses in the early Universe, in order to account
for the G dwarf problem, the large heavy element abundance and high
temperature in galactic cluster gas, and the high luminosities of
distant galaxies. Variations like this would be more easily explained by
an IMF model that depends on the thermal Jeans mass.

The thermal Jeans mass, which contains the combination of
parameters $T^2/P^{1/2}$, is approximately
constant in normal regions of star
formation. This is because the numerator in this expression is
approximately proportional to the cooling rate per unit mass in
molecular clouds (which scales about as $T^2-T^3$ 
-- see Neufeld, Lepp, \& Melnick 1995), and the denominator is
approximately proportional to the heating rate per unit mass from
starlight and cosmic rays in typically active disks. The starlight and
cosmic ray intensities scale with the background column density of
stars, and the pressure in the midplane of the disk scales with the
square of this column density. Thus the square root of pressure goes
with the column density of background stars. As long as heating equals
cooling and the mass-to-light ratio in a galactic disk
is about constant, and as long as the factor by which
star-forming clouds have a higher pressure than
the ambient pressure is about constant,
the thermal Jeans mass is about the same in all 
dense cloud regions.
If the mass-to-light ratio goes down, then the thermal Jeans
mass can go up. Perhaps this occurs in starburst regions.
Conversely, if the mass-to-light ratio is abnormally high, then
the thermal Jeans mass can go down.

An example of the latter situation might arise in
the inner regions of M31. There the molecular cloud heating
rate is low and the cloud temperature is close to $3$K, instead of the
usual 10K (Allen et al. 1995; Loinard \& Allen 1998).
These clouds also exist in the part of the
disk where the stellar column density is high in old stars, so the
interstellar pressure is not particularly low. As a result,
the thermal Jeans mass can be lower in ultracold clouds than in normal clouds,
possibly as
low as $0.01$ M$_\odot$ instead of 0.3 M$_\odot$
(Elmegreen 1999c). For this reason, a
significant population of Brown Dwarf stars might be present in
ultracold molecular clouds. If they are found, then the model based on the 
thermal
Jeans mass would be preferred over the model based on the deuterium
burning limit.

The thermal Jeans model is preferred also if a reasonably high fraction,
say $>10$\%, of all the material in a collapsing cloud piece gets into
a star. This leaves a lot of mass for wind expulsion and disk erosion,
but it also implies that the star mass depends somewhat on the mass of
the cloud piece in which it forms. In that case, wind-limitations to the
stellar mass would not be very important, causing only a factor of 2--10
variation in the ratio of star mass to cloud mass. Most of the mass
variation along the IMF, which spans a factor of $\sim10^3$ in mass,
would then have to come from something else, and the mass of the pre-stellar
cloud piece is a likely place.

Another observation that could help distinguish between possible
origins for the characteristic stellar mass is the discovery of
powerful pre-main sequence winds from extremely low-mass Brown Dwarfs,
i.e., stars too small to ignite even deuterium.  If pre-main sequence
contraction energy alone is enough to start a wind, then deuterium
burning would not be relevant to the limitation of stellar mass.

There is some evidence already that the mass function for dense cloud cores
containing about a solar mass is similar to the IMF (Motte,
Andr\'e, \& Neri 1998; Testi \& Sargent 1998).
This is the type of observation that could clarify the origin of the
characteristic mass for star formation.

\subsection{Top-Heavy IMFs in Starburst Regions}
\label{sect:sb}

There has been considerable discussion about a shift in the IMF
towards proportionally more high mass stars in starburst regions,
although many of the initial reports are now being questioned.
The original motivation for this idea was the observation that
the luminosity of the starburst was so high, given the total
mass from the rotation curve, that there could not be a normal
proportion of high and low mass stars but only an excess of
high mass stars. Now, more
detailed modeling, and in the case of M82, a lower extinction
correction (Devereux 1989, Satyapal et al. 1997), makes the
stellar luminosity seem about right for the mass. A summary
of these observations is in Table 4. In addition, a top-heavy IMF
would produce too much oxygen in proportion to other elements
(Wang \& Silk 1993), and the aging population of stars would be
too red (Charlot et al. 1993).

\begin{table}
\caption{Key IMF Observations in Starburst Galaxies}
\begin{tabular}{ll}
\tableline
Shifts to high mass&Rieke et al.  1980; Kronberg, Biermann \& Schwab\\
& 1985; Wright et al.  1988; Telesco 1988; Doane\\
& \& Matthews 1993; Doyon, Joseph, \& Wright 1994\\
& Rieke et al.  1993;
Smith et al.  1995, 1998\\
& Shier et al. 1996\\
Normal, $x\sim1-1.5$&Devereux 1989; Schaerer 1996; Satyapal et al.\\
& 1997; Calzetti 1997\\
\tableline
\tableline
\end{tabular}
\end{table}

Considering the basic form of the IMF, which is a power
law with a lower cutoff or flattening at some
characteristic mass, one can easily envision variations
that lead to top-heavy IMFs as a result of an upward shift in the
characteristic mass.  A predicted downward shift leading to an excess of
Brown Dwarfs was mentioned in a previous section.
The upward shift would come in the same way, but from an increase rather
than a decrease in the value of $T^2/P^{1/2}$.  It is more difficult
to envision a top heavy IMF that results from a decrease
in the slope of the power law part, because the very existence
of a power law suggests a scale-free process, which means that it
is essentially free of dependence on physical parameters.
Power law mass distributions often result from geometric (e.g., fractal)
or self-regulatory (e.g., equilibrium coalescence) effects instead.

The IMF model in Elmegreen (1997), in which the power law part comes
from a weighted selection of clump pieces in a hierarchically structured
cloud and the low mass cutoff comes from the thermal
Jeans mass, gets a simple shift in the whole IMF towards higher mass,
with a constant slope in the power-law part, as $T^2/P^{-1/2}$
increases. A computer simulation showing this result was in that paper.

An amazing thing about the IMF is that the characteristic mass at the
low end, where the flattening occurs, appears to be nearly
constant from region to
region. As discussed above, this may simply reflect equilibrium thermal
conditions with varying $T$ and $P$ but constant $T^2/P^{1/2}$, or it
may reflect a constant wind-limited mass at the threshold of deuterium
burning. The upward shift for starbursts, if real, provides a
good test for the models. It is easier to increase $T^2/P^{1/2}$ in warm
regions at slightly elevated pressures than to affect the deuterium
burning limit, which would seem to be independent of environment. Thus
the exact form of the IMF in starburst conditions is extremely important
for the models. In this respect, the reported slight upward shift in the
characteristic mass for the 30 Dor cluster in the LMC (Nota et al. 1998)
is noteworthy. This is the closest starburst-like region, and therefore
the most promising for providing a firm observation of the IMF from
direct star counting.  Unfortunately, this cluster could
suffer from mass segregation effects as in other
clusters, in which case the upward shift
would appear only in the nuclear region. 

The discussion about starburst IMFs begs the question of whether
there is an upper limit to the mass of a star that can form.
No such upper limit has been found yet. That is, the upper limit in any
particular region just keeps increasing as the total stellar mass increases,
as expected for random star formation (see theory in
Elmegreen 1983, 1997, and observations in Massey \& Hunter 1998).
Yet there would seem to
come a time where this stellar mass increase would have to stop. After all,
if we scale the $1/x$ power law relation between the maximum star mass and
total star mass to all of the young stellar mass in the galaxy, with
an age less than the $\sim2$ million year lifetime of a massive star,
then the total young stellar mass is $\sim10^7$ M$_\odot$ and the
expected maximum stellar mass is
\begin{equation}M_{max}\sim 50 \left({{10^7\;{\rm M}_\odot}\over
{10^{4.5}\;{\rm M}_\odot}}\right)^{1/1.35}\;{\rm M}_\odot\sim3600\;{\rm M}_\odot
.
\end{equation}
Here we have normalized this power law relation to the maximum mass ($\sim50$ M$
_\odot$)
and total mass ($\sim10^{4.5}$ M$_\odot$) in the Orion OB association.
The result is very inaccurate, of course, but the lack of Galactic stars
containing several thousand solar masses suggests that there is an upper
mass cutoff.

An alternative explanation for the lack of thousand-M$_\odot$ stars is
that each star-forming region is independent, so the total stellar mass
used in the above equation is the maximum stellar mass in the largest
region of star formation, rather than the maximum for all regions in the
Galaxy. In that case, the numerator in the above expression should be
$\sim10^{5.5}$ M$_\odot$ for the largest star complexes forming in $10^7$
M$_\odot$ spiral arm clouds, and $M_{max}\sim200$ M$_\odot$, which may
be possible a few places in the Galaxy.  If such stars are found, then
there may be no maximum mass based on physical principles, only one
based on sampling statistics.

\section{Peculiarities with Massive Stars: central concentration and
late appearances in clusters, and a preference for massive clouds}
\label{sect:center}

Most massive stars form in giant molecular clouds in OB associations,
and not in small clouds like Taurus, which seem to contain only low mass
stars (Larson 1982; Myers \& Fuller 1993). Massive stars also form
relatively late in the evolution of a star cluster, after many low mass
stars have already formed (Herbig 1962; Iben \& Talbot 1966; Herbst \&
Miller 1982; Adams, Strom \& Strom 1983).

There have been several attempts to explain the correspondence between
extreme star mass and cloud mass as a consequence of different
mechanisms for star formation or different physical conditions in large
and small clouds (Larson 1992; Khersonsky 1997), however observations
like this are expected from random star formation alone (Elmegreen 1983;
Walter \& Boyd 1991; Massey \& Hunter 1998), so the need for any special
theory is not compelling.

If stars form randomly in all clouds, with stellar masses in
the proportion given by a normal IMF, then statistical effects will make
the massive stars, which are relatively rare, more likely to appear
after there are $100-1000$ M$_\odot$
of other stars already (Elmegreen 1983; Schroeder
\& Comins 1988). This means that massive stars tend to show up only
in massive clouds, and when they do, they are relatively late
compared to the more common low mass stars. Simulations of this effect
are in Elmegreen (1999a).
Note that the average time of appearance of a 
star with a particular mass is still independent of that
mass in this statistical interpretation, so if there is 
a {\it systematic} bias toward a late appearance 
of high mass stars, then some physical process for this would be
required. Stahler (1995) suggested, however,
that even the proposed examples of such bias probably
have other interpretations, so the 
entire effect could be just statistical. 

Another peculiar observation of massive stars is that they tend to
appear near the centers of star clusters, surrounded by the lower mass
stars (see reviews in Elmegreen et al. 1999; Testi, Palla, \& Natta
1998). This peculiar distribution for massive stars has been observed
using color gradients in 12 clusters (Sagar \& Bhatt 1989), and from the
steepening of the IMF with radius in several clusters (Pandey, Mahra, \&
Sagar 1992), including Tr 14 (Vazquez et al. 1996), the Trapezium in
Orion (Jones \& Walker 1988; Hillenbrand 1997; Hillenbrand \& Hartmann
1998), and, in the LMC, NGC 2157 (Fischer et al. 1998), SL 666, and NGC
2098 (Kontizas et al. 1998).

The usual explanation for this effect is that massive stars sink to the
center of a cluster during dynamical relaxation, but several clusters
seem too young for this to have happened (Bonnell \& Davies 1998),
including Orion Trapezium (Hillenbrand \& Hartmann 1998). In that case,
the high mass stars had to have been born near the cluster centers,
perhaps because the most massive clumps were closer to the center at the
time the massive stars were born in them. There are other explanations
too. The stars near the center could have accreted gas faster and ended
up more massive (Larson 1978, 1982; Zinnecker 1982; Bonnell et al.
1997); they or their predecessor clumps could have coalesced more
(Larson 1990; Zinnecker et al. 1993; Stahler, Palla, \& Ho 1999;
Bonnell, Bate, \& Zinnecker 1998), or the most massive stars and clumps
forming anywhere could have migrated to the center faster because of a
greater gas drag (Larson 1990, 1991; Gorti \& Bhatt 1995, 1996;
Saiyadpour, Deiss, \& Kegel 1997). 
A problem with most of these models is that
they are inconsistent with the
observation that the IMF is nearly
independent of cluster
density (Sect. 2.3). Another
model without this problem suggests that the central location of
the most massive stars is from the central location of the most
massive cloud pieces, which is expected for a hierarchical cloud (Elmegreen
1999a). 

\section{Evolution of the IMF}

The discussion above suggests that the IMF has been somewhat constant in
time and place, except possibly for an upward shift in the mass at the
IMF peak for starburst regions (Sect. 2.5). There was also a suggestion
that the IMF was shifted towards higher mass in the early Universe
(Larson 1998), although very old stars and old intergalactic gas seem to
show evidence for a normal IMF (Table 1).

The direct observations of normal star-forming regions point to
a {\it universal} IMF, with deviations perhaps only from statistical
fluctuations in small samples and from mass
segregation in clusters. The observations of regions with extremely low
star-forming activity suggest a shift towards lower masses, either with
a steeper IMF (as observed by Massey et al. 1995) or, possibly, a downward
shift in the peak (as predicted by Elmegreen 1999c). Observations of
regions with extremely high star-forming activity suggest an analogous
shift towards higher masses, possibly as a result of an upward shift in the
peak. 

If the mass at the peak of the IMF can really change with star formation
activity, possibly as a result of changes in the ratio $T^2/P^{1/2}$,
which is in the thermal Jeans mass, then there are
several important implications. First, the ratio $T^2/P^{1/2}$ depends
roughly on the light-to-mass ratio in a galaxy disk, because the
numerator is proportional to the cooling, and therefore heating rate in
molecular clouds, and the denominator is proportional to the local mass
column density (Sect. 2.4). This means that if the light-to-mass ratio
is high, the peak in the IMF can shift towards higher masses, and
vice versa. Now it follows from the Schmidt law, which has a star
formation rate proportional to average density to some power greater
than unity (e.g., Kennicutt 1998), that the gas consumption rate in a
star-forming region increases with higher density, and 
the luminosity-to-mass ratio for luminous young stars increases too.
If the peak in the IMF increases along with the higher L/M ratio, then we get
the interesting result that the IMF peak
increases with 
the gas consumption rate (Elmegreen 1999a). We might also have
a higher efficiency of star formation in such a region,
because of the generally greater self-binding of clouds in
high pressure or high velocity-dispersion gas (Elmegreen,
Kaufman, \& Thomasson 1993).  This circumstance could then explain 
why some starburst regions have all three of these
peculiarities at the same time (see review in Telesco 1988).

What happened in the early Universe is more difficult to assess. Although
the temperature was higher from the cosmic microwave background, in
proportion to $(1+z)$, the average density of the Universe was higher
too, in the proportion $(1+z)^3$, and the pressure, which is a product
of density and temperature, was higher by $(1+z)^4$. Thus the ratio of
$T^2/P^{1/2}$ in the thermal Jeans mass was independent of $z$. However,
$T$ and $P$ variations in newly forming galaxies should dominate these
average $z$ variations, and the thermal Jeans mass could have gone either way.
If the earliest stars formed in cool high-pressure shocks, then perhaps the
thermal Jeans mass was lower than it is today, producing Brown
Dwarfs. On the other hand, if the temperature was higher because of an
inability to cool without metals, then the Jeans mass could have been
higher. The observation of nearly normal abundances in Ly $\alpha$
forest lines and the intercluster medium (Table 1) suggest that this
characteristic mass probably did not vary much in the early Universe. 

\section{Summary}

The IMF is a power law at intermediate to high mass, with a flattening
on a log-log plot at low mass. The mass at which the flattening occurs
is the only characteristic mass that has been clearly observed for
star formation, and is therefore an important 
indicator of physical processes that depend on scale. Examples might
be the thermal Jeans mass or the minimum mass for deuterium burning
and stellar winds, both of which have about the right value. Methods
to distinguish between these two possibilities were discussed
in Section 2.4.  The power-law part of the IMF may not indicate
specific physical processes, but be more of a remnant from the observed
scale-free geometry of pre-stellar clouds.   Random sampling models
for such geometries reproduce essentially all of the IMF properties
with very little sensitivity to free parameters. In this case, much of the
physics of the star formation process may be unrecoverable from
the power-law part of the IMF alone. 

The lack of any obvious dependence of the IMF on cluster density places
strong constraints on the physical processes that might be involved
(Sect. 2.3).  The steep IMF in the extreme field (Sect. 2.2), as well
as other systematic variations in the IMF, such as the concentration of
massive stars in cluster cores (Sect. 3) and the shift in the IMF
towards higher masses in starburst regions (Sect. 2.5), all suggest
specific physical differences in the properties of star-forming regions
and perhaps in the mechanisms of star formation too.
Differences in the IMF
from place to place and time to time may eventually tell us more about star
formation than any single IMF, which may have washed out any such 
details in the averaging process.

\end{document}